\documentclass[twocolumn]{aastex61}
\usepackage{amsmath,amstext}
\usepackage{apjfonts} 
\usepackage{afterpage}
\usepackage{color}
\usepackage{multirow,bigdelim}
\usepackage[hang,flushmargin]{footmisc}
\renewcommand{\arraystretch}{1.5}
\usepackage{soul}
\setstcolor{red}

\usepackage{hyperref}
\usepackage{url}
\usepackage{booktabs}

\received{--}
\revised{--}
\accepted{--}

\shorttitle{Two years of non-thermal emission from GW\,170817}
\shortauthors{Hajela et al.}
\begin{document}

\title{Two years of non-thermal emission from the binary neutron star merger GW\,170817: rapid fading of the jet afterglow and first constraints on the kilonova fastest ejecta}

\author{{A. ~Hajela}\altaffilmark{1}, {R. ~Margutti}\altaffilmark{1,2}, {K. D. ~Alexander}\altaffilmark{1,3}, {A. ~Kathirgamaraju}\altaffilmark{4,5}, {A. ~Baldeschi}\altaffilmark{1}, {C. ~Guidorzi}\altaffilmark{6}, {D. ~Giannios}\altaffilmark{4}, {W. ~Fong}\altaffilmark{1}, {Y. ~Wu}\altaffilmark{7}, {A. ~MacFadyen}\altaffilmark{7}, 
{A. Paggi}\altaffilmark{8,9}, {E. ~Berger}\altaffilmark{10}, {P. K. ~Blanchard}\altaffilmark{1}, {R. ~Chornock}\altaffilmark{11}, {D. L. ~Coppejans}\altaffilmark{1}, {P. S. ~Cowperthwaite}\altaffilmark{12,13}, {T. ~Eftekhari}\altaffilmark{10}, {S. ~Gomez}\altaffilmark{10}, {G. ~Hosseinzadeh}\altaffilmark{10}, {T. ~Laskar}\altaffilmark{14}, {B. D. ~Metzger}\altaffilmark{15}, {M. ~Nicholl}\altaffilmark{16,17}, {K. ~Paterson}\altaffilmark{1}, {D. ~Radice}\altaffilmark{18,19}, {L. ~Sironi}\altaffilmark{20}, {G. ~Terreran}\altaffilmark{1}, {V. A. ~Villar}\altaffilmark{10}, {P. K. G. ~Williams}\altaffilmark{10,21}, {X. Xie}\altaffilmark{22}, and {J. ~Zrake}\altaffilmark{15}} 

\altaffiltext{1}{Center for Interdisciplinary Exploration and Research in Astrophysics and Department of Physics and Astronomy, Northwestern University, 2145 Sheridan Road, Evanston, IL 60208-3112, USA}
\altaffiltext{2}{CIFAR Azrieli Global Scholar, Gravity \& the Extreme Universe Program, 2019}
\altaffiltext{3}{NASA Einstein Fellow}
\altaffiltext{4}{Department of Physics and Astronomy, Purdue University, 525 Northwestern Avenue, West Lafayette, IN 47907-2036, USA}
\altaffiltext{5}{Department of Astronomy and Theoretical Astrophysics Center, University of California Berkeley, Berkeley, CA 94720}
\altaffiltext{6}{Dipartimento di Fisica, Universit\`{a} Ferrara, Via Paradiso 12, I-44100 Ferrara, Italy}
\altaffiltext{7}{Center for Cosmology and Particle Physics, New York University, 726 Broadway New York, NY 10003, USA}
\altaffiltext{8}{INAF - Osservatorio Astrofisico di Torino, via Osservatorio 20, 10025 Pino Torinese, Italy}
\altaffiltext{9}{INFN - Istituto Nazionale di Fisica Nucleare, Sezione di Torino, via Pietro Giuria 1, I-10125 Turin, Italy}
\altaffiltext{10}{Center for Astrophysics | Harvard \& Smithsonian, 60 Garden Street, Cambridge, MA 02138-1516, USA}
\altaffiltext{11}{Astrophysical Institute, Department of Physics and Astronomy, 251B Clippinger Lab, Ohio University, Athens, OH 45701, USA}
\altaffiltext{12}{Observatories of the Carnegie Institute for Science, 813 Santa Barbara Street, Pasadena, CA 91101-1232, USA}
\altaffiltext{13}{NASA Hubble Fellow}
\altaffiltext{14}{Department of Physics, University of Bath, Claverton Down, Bath, BA2 7AY, UK}
\altaffiltext{15}{Department of Physics and Columbia Astrophysics Laboratory, Columbia University, New York, NY 10027, USA}
\altaffiltext{16}{Institute for Astronomy, University of Edinburgh, Royal Observatory, Blackford Hill, EH9 3HJ, UK}
\altaffiltext{17}{Birmingham Institute for Gravitational Wave Astronomy and School of Physics and Astronomy, University of Birmingham, Birmingham B15 2TT, UK}
\altaffiltext{18}{Department of Physics, The Pennsylvania State University, University Park, PA 16802, USA}
\altaffiltext{19}{Department of Astronomy \& Astrophysics, The Pennsylvania State University, University Park, PA 16802, USA}
\altaffiltext{20}{Department of Astronomy and Columbia Astrophysics Laboratory, Columbia University, New York, NY 10027, USA}
\altaffiltext{21}{American Astronomical Society, 1667 K Street NW, Suite 800, Washington, DC 20006-1681, USA}
\altaffiltext{22}{Mathematical Sciences and STAG Research Centre, University of Southampton, Southampton SO17 1BJ, United Kingdom}

\begin{abstract}
 We present Chandra and VLA observations of GW\,170817 at $\sim521-743$ days post merger, and a homogeneous analysis of the entire Chandra dataset. We find that the late-time non-thermal emission follows the expected evolution of an off-axis relativistic jet, with a steep temporal decay $F_{\nu}\propto t^{-1.95\pm0.15}$ and power-law spectrum $F_{\nu}\propto \nu^{-0.575\pm0.007}$. We present a new method to constrain the merger environment density based on diffuse X-ray emission from hot plasma in the host galaxy and find $n\le 9.6 \times 10^{-3}\,\rm{cm^{-3}}$. This measurement is independent from inferences based on  jet afterglow modeling and allows us to partially solve for model degeneracies. The updated best-fitting model parameters with this density constraint are a fireball kinetic energy $E_0 = 1.5_{-1.1}^{+3.6}\times 10^{49}\,\rm{erg}$ ($E_{iso}= 2.1_{-1.5}^{+6.4}\times10^{52}\, \rm{erg}$), jet opening angle $\theta_{0}= 5.9^{+1.0}_{-0.7}\,\rm{deg}$ with characteristic Lorentz factor $\Gamma_j = 163_{-43}^{+23}$, expanding in a low-density medium with $n_0 = 2.5_{-1.9}^{+4.1} \times 10^{-3}\, \rm{cm^{-3}}$ and viewed $\theta_{obs} = 30.4^{+4.0}_{-3.4}\, \rm{deg}$  off-axis. The synchrotron emission originates from a power-law distribution of electrons with index $p=2.15^{+0.01}_{-0.02}$. The shock microphysics parameters are constrained to $\epsilon_{\rm{e}} = 0.18_{-0.13}^{+0.30}$ and $\epsilon_{\rm{B}}=2.3_{-2.2}^{+16.0} \times 10^{-3}$. Furthermore, we investigate the presence of X-ray flares and find no statistically significant evidence of $\ge2.5\sigma$ of temporal variability at any time. Finally, we use our observations to constrain the properties of synchrotron emission from the deceleration of the fastest kilonova ejecta with energy $E_k^{KN}\propto (\Gamma\beta)^{-\alpha}$ into the environment, finding that shallow stratification indexes $\alpha\le6$ are disfavored. Future radio and X-ray observations will refine our inferences on the fastest kilonova ejecta properties.
\end{abstract}

\section{Introduction}

Multi-messenger observations of the binary neutron star (BNS) merger event GW\,170817 ushered us into a new era of systematic exploration of our universe with gravitational waves and electromagnetic emission \citep{abbott2017gw170817,abbott2017multi}.  Light from GW\,170817 has been detected across the electromagnetic spectrum, from the $\gamma$-rays to the radio wavelengths (\citealt{Savchenko+2017}, \citealt{blanchard2017electromagnetic}, \citealt{SwopeDiscovery}, \citealt{valenti2017discovery}, \citealt{chornock2017electromagnetic}, \citealt{cowperthwaite2017electromagnetic}, \citealt{nicholl2017electromagnetic}, \citealt{Fong+2017}, \citealt{margutti2017electromagnetic}, \citealt{alexander2017electromagnetic}, \citealt{haggard2017chandra}, \citealt{hallinan2017counterpart}, \citealt{kasliwal2017illuminating}, \citealt{troja2017x}, \citealt{Dobie+2018}, \citealt{lyman2018optical}, \citealt{nynka2018}, \citealt{ruan2018x}, \citealt{Margutti+2018}, \citealt{Alexander+2018}). While the radiation powering the thermal emission from the kilonova (KN) associated to the BNS merger peaked at $\delta t<12$ days (e.g. \citealt{villar2017combined}, their Fig. 1) and  faded below the detection threshold of current instrumentation at $\delta t\sim 70$ days (with the latest detections in the NIR, \citealt{Villar+2018,kasliwalspitzer2019}), the non-thermal emission from the off-axis structured relativistic jet is longer lived.

Here we present deep X-ray and radio observations of the non-thermal emission from GW\,170817 covering the period $\delta t\sim521-743$ days with the  Chandra X-ray Observatory (CXO) and the Karl G. Jansky Very Large Array (VLA), together with a comprehensive re-analysis of the entire CXO data set. These observations allow us to refine previous inferences on the physical properties of the relativistic outflow launched by the BNS merger, and the density of the  environment where the outflow is expanding \citep{Alexander+2018,Davanzo2018,Dobie+2018,Granot+2018,Hotokezaka18,Margutti+2018,Mooley3+2018,Mooley1+2018,Mooley2+2018,Lazzati2018,Fong19,Ghirlanda2019,Lamb+2019,troja+2018,Troja2019}. Finally, we use these observations to put the first constraints on the properties of non-thermal synchrotron emission from the deceleration of the KN fastest ejecta (i.e. the KN afterglow, e.g. \citealt{nakarpiran2011,Kathirgamaraju19}).

This paper is organized as follows. We present the analysis of two recent CXO observations at $\delta t \sim582$ and $\delta t \sim743$  days in  \S\ref{Sec:CXOdata}, together with  a homogeneous temporal and spectral re-analysis of the entire CXO data set acquired in two years of monitoring of GW\,170817. New VLA observations of GW\,170817 at $\delta t>500$ days are presented in \S\ref{sec:radio}. \S\ref{SubSec:afterglow} is dedicated to the broad-band modeling of the non-thermal emission from GW\,170817 within the boosted fireball framework of \cite{WuMacfadyen2018}. Constraints on the KN afterglow and the physical properties of the KN fastest ejecta are derived in \S\ref{SubSec:KNejecta}. Conclusions are drawn in \S\ref{Sec:Conc}.

All times are measured with respect to the time of the gravitational-wave trigger, which is August 17th 2017 12:41:04 UT \citep{abbott2017gw170817}. Uncertainties are provided at the $1\,\sigma$ confidence level (c.l.) and upper limits at the $3\,\sigma$ c.l. unless otherwise stated. We adopt the  luminosity distance of NGC\,4993, the host galaxy of GW\,170817,  $d=40.7$  Mpc inferred by \cite{Cantiello2018}. 
\section{X-ray Data Analysis}
\label{Sec:CXOdata}
The Chandra X-ray Observatory (CXO) started observing GW\,170817 on 2017 August 19 ($\delta t \sim$2 days after the merger). Here we use a uniform framework for data reduction to perform a temporal and spectral analysis of new observations acquired at $\delta t \sim 580-740$ days, and a re-analysis of the entire Chandra data set spanning $\delta t \sim 2 - 356$ days after the merger.  This is fundamental to our analysis and enables us to consistently compare the fluxes, measure the ambient density of the merger environment, reliably search for temporal variability, and model the afterglow.
The  total exposure time across all observations is $\sim$ 731 ks (Table \ref{tab:wavdetecttab1}). The CXO data set acquired at  $\delta t \sim$ 582 and $\sim$ 743 days is presented here for the first time. Previous CXO observations of GW\,170817 have been presented by \cite{margutti2017electromagnetic,haggard2017chandra,troja2017x,Alexander+2018,nynka2018,Margutti+2018,troja2018x,ruan2018x,Pooley2018,Piro+2019}. 
Our analysis consistently accounts for the low count statistics of the Chandra observations of GW\,170817 to accurately determine the model parameters and their uncertainties, as described in \cite{margutti2017electromagnetic,Alexander+2018,Margutti+2018}.  We show the XMM measurements from \cite{Davanzo2018,Piro+2019} in Fig. \ref{fig:Xray} but we do not include these data in our modeling below to minimize the impact of systematic effects arising from, for instance, variability of the central AGN confused with GW~170817 in the XMM PSF.

\subsection{X-ray Temporal Analysis of  GW\,170817}
\label{SubSec:temp}

 We homogeneously reduced the entire CXO data set acquired at $\delta t \sim 2-743$ days since the merger following the steps below. We  reprocessed all the observations with the \texttt{repro} task within 
 \texttt{CIAO} (v4.11, \citealt{Fruscione2006})
applying standard ACIS data filtering and using the latest calibration database (\texttt{CALDB}, v4.8.3).  We performed blind point-source detection with \texttt{wavdetect} on individual observation IDs. The results are reported in Table \ref{tab:wavdetecttab1}. An X-ray source is blindly detected with \texttt{wavdetect} at a location consistent with GW\,170817 in all observations acquired at $9.2\le\delta t<360$ days, with inferred 0.5--8 keV net count-rates reported in Table \ref{tab:wavdetecttab1}. No X-ray emission from GW\,170817 is detected at $\delta t\sim2.3$ days and our results are consistent with the earlier report by \cite{margutti2017electromagnetic}. The X-ray counterpart of GW\,170817 is blindly detected with a very low significance ($<3\,\sigma$) in the individual observations acquired at the epochs corresponding to $\delta t\sim 581 - 743$ days (IDs $21322$, $22157$, $22158$, $21372$,  $22736$, $22736$; PIs: Margutti, Fong, Troja; programs $20500299$, $20500691$). 
 However, we note that 
 GW\,170817 is blindly detected with significance $\ge3\,\sigma$ when 
 observations acquired around the same time are merged (grouped IDs in Table \ref{tab:wavdetecttab1}). 

Motivated by the claim of significant temporal variability around $\sim160$ days by \cite{Piro+2019}, we searched for short time-scale variability within each observation ID  and for observations acquired within $\delta t/t\le0.03$ (i.e. grouped IDs in Table \ref{tab:wavdetecttab1}) by applying a  multinomial test to the observed photon counts. The null hypothesis that we want to test is that of a constant source count-rate in a time interval $\Delta t_{tot}$. We thus assigned to each time interval a probability proportional to the effective exposure time $\Delta t_k$ within $\Delta t_{tot}$, and computed the log-likelihood of the observed photon counts with respect to a multinomial distribution with $n=N_{tot}$ (where $n$ is the number of trials and $N_{tot}$ is the total number of observed photons in $\Delta t_{tot}$). 
 We then generated 
$10^4$ realizations of $N_{tot}$ events distributed among  $\Delta t_k$ following a multinomial distribution with probabilities defined as above. For each $\Delta t_{tot}$, the statistical significance of the evidence of a departure from our null hypothesis is quantified by the fraction of synthetic data sets that showed a log-likelihood value at least as extreme as the one observed. 
We applied the multinomial test to each observation ID and to grouped IDs in Table \ref{tab:wavdetecttab1}. For single IDs, $\Delta t_{tot}$ is defined by the start and end of the CXO observations and we divided $\Delta t_{tot}$ into two halves, $\Delta t_{1}$ and $\Delta t_{2}$. For grouped IDs, $\Delta t_{tot}$ encompasses the time interval defined by the beginning and end time of the first and last observation, respectively, and the values of $\Delta t_{k}$ are naturally defined as the exposure times of each ID. 
 
We find no evidence for departures from our null hypothesis in the entire sample of CXO observations of GW\,170817, with a statistical significance of short time-scale variability of the X-ray emission from GW\,170817  of $\le2.5\sigma$ (Gaussian equivalent). 
In particular, our results do not confirm the claim of temporal variability at the level of $3.3\,\sigma$ in the time interval $\Delta t_{tot}=153-164$ days by \cite{Piro+2019}. By applying the same method as in \cite{Piro+2019} we find that we can reproduce their results only for their particular choice of time intervals ($\Delta t=153.4-157.2$ days  vs. $\Delta t=159.8-163.8$ days, without considering data acquired in CXO ID 20937) and only if we do not account for the number of trials.\footnote{Excluding the central portion of the data as in \cite{Piro+2019}, but allowing for a random selection of the initial and final time intervals to compare, leads to the conclusion that only $\sim0.4\%$ of blind choices 
would lead to a claim of temporal variability as significant as $\sim3.3\sigma$ (see detailed discussion in Appendix \ref{App:variability} and Fig. \ref{fig:variability}).} Properly accounting for the trials with the test above leads to a reduced statistical evidence for temporal variability in this time interval   
of $1.8\,\sigma$. We thus conclude that there is no statistical evidence for short-term variability in the X-ray afterglow of GW\,170817 and that the current CXO data set does not quantitatively support the notion of an X-ray flare from a surviving magnetar remnant at $\delta t\sim 160$ days (\citealt{Piro+2019}).

\subsection{X-ray Spectral analysis of  GW\,170817}
\label{SubSec:spec}

\begin{figure*}[!t]
\begin{center}
\hspace*{-0.1in}
{\includegraphics[scale=0.6]{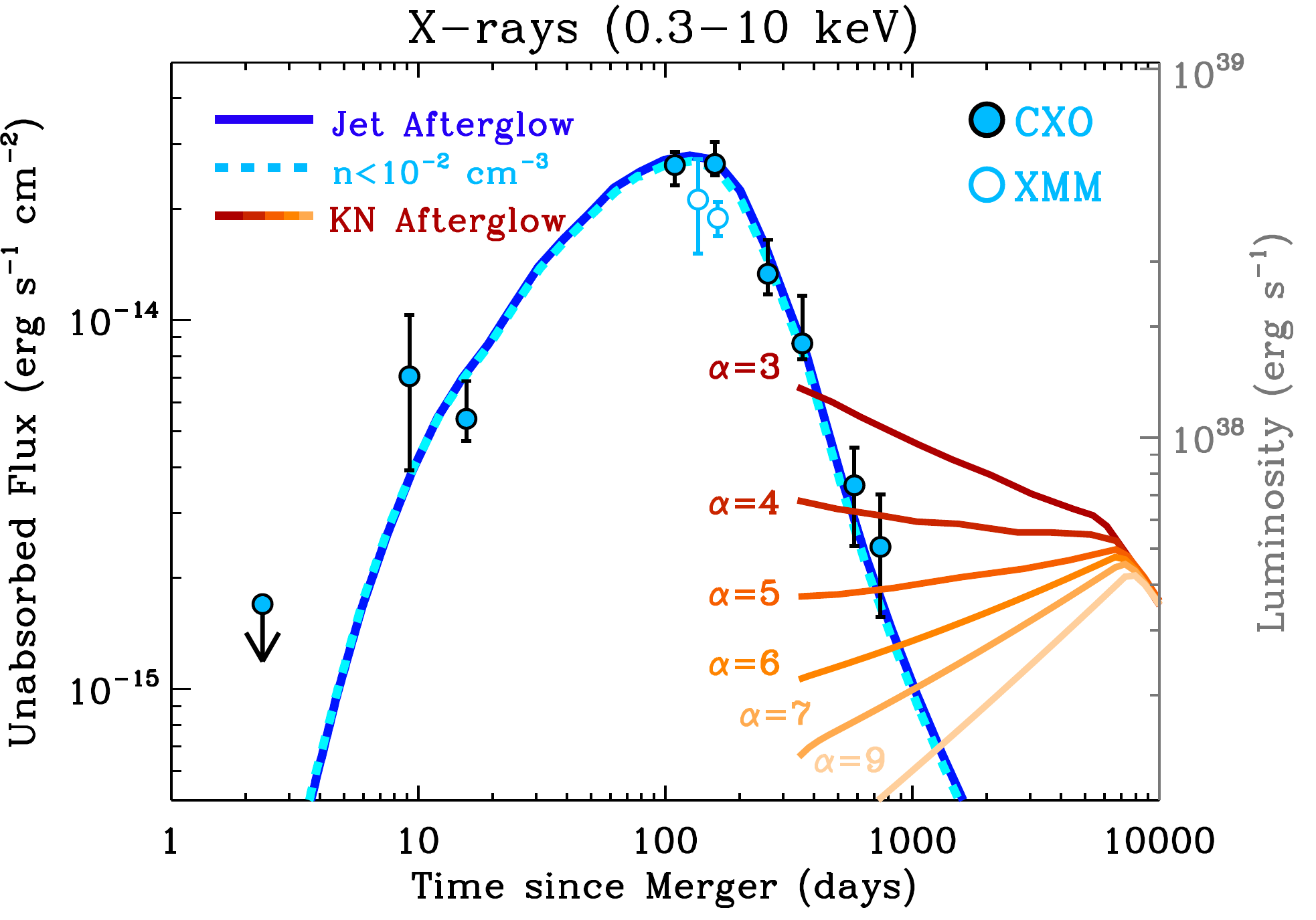}
\includegraphics[scale=0.61]{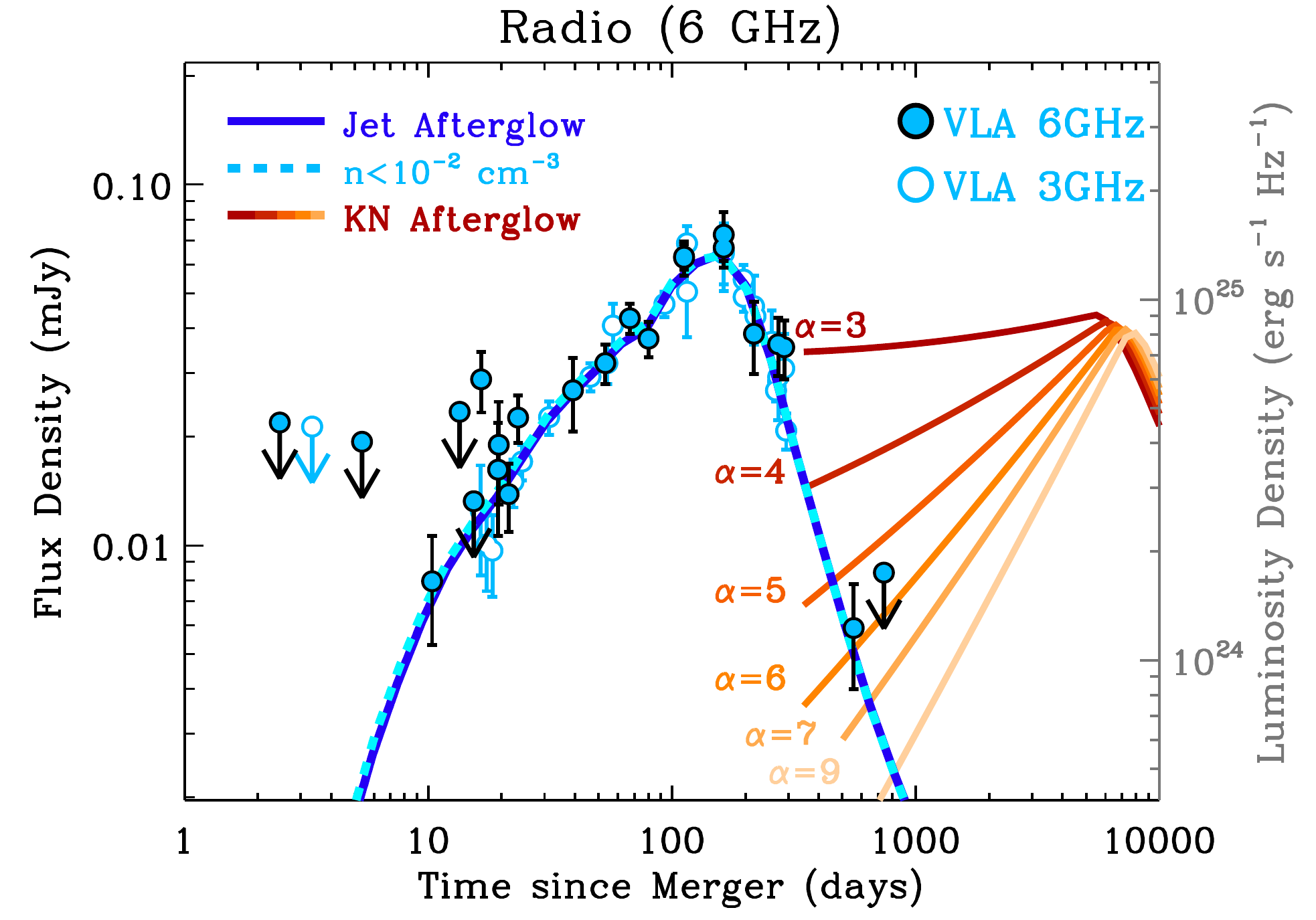}}
\caption{X-ray (upper panel) and radio (lower panel) emission from GW\,170817 in the first $\sim743$ days since merger as constrained by the CXO (this work), XMM-Newton (\citealt{Davanzo2018,Piro+2019}) and the most recent VLA observations (this work) merged with previous VLA observations (\citealt{Alexander+2017}, \citealt{hallinan2017counterpart},\citealt{Mooley1+2018}, \citealt{Margutti+2018},\citealt{Alexander+2018}, \citealt{Mooley2+2018}).
We plot the VLA 6 GHz data (filled circles) and the 3 GHz data (empty circles) scaled at 6 GHz using an $F_{\nu}\propto \nu^{-0.6}$ spectrum.
The broad-band emission continues to be well modeled by a structured off-axis jet (solid blue line) with best fitting energy $E_0\sim 2 \times 10^{50}$ erg, $\theta_{obs}\sim 33^{\circ}$, $\theta_{0}\sim 7^{\circ}$ propagating into a medium with density $n\sim 0.07$ cm$^{-3}$ (\S\ref{SubSec:afterglow}, Fig. \ref{fig:contourcorner}). Dashed light-blue lines: best fitting structured jet model 
for $n< 9.6\times10^{-3}\,\rm{cm^{-3}}$ as derived in \S\ref{SubSec:newmethod}, which leads to $E_{0}\sim 1.5 \times 10^{49}$ erg, $\theta_{obs}\sim 30^{\circ}$, $\theta_{0}\sim 6^{\circ}$ (\S\ref{SubSec:afterglow}, Fig. \ref{fig:contourdensUL}).
Thick red-to-orange lines: expected emission originating from the deceleration of the KN ejecta into the environment (i.e. the KN afterglow). We adopt the parametrization by \cite{Kathirgamaraju19} and show the expected KN afterglow emission for a set of representative values of the stratification index $\alpha=3,4,5,6,7,8,9$ of the KN ejecta kinetic energy  $E_{k}^{KN}(>\Gamma\beta)\propto (\Gamma \beta)^{-\alpha}$,  and for fiducial values of the microphysical parameters $\epsilon_B=10^{-3}$, $\epsilon_e=0.1$. We further adopt an environment density $n=0.01\,\rm{cm^{-3}}$ (the largest value allowed by our modeling of the diffuse X-ray emission, 
and a KN outflow with minimum velocity $v_0\sim0.3$c and total energy $\sim10^{51}\,\rm{erg}$, as found from the modeling of the UV-optical-NIR KN emission, which is sensitive to the slower moving ejecta that carries the bulk of the KN kinetic energy (e.g. \citealt{villar2017combined}). Current observations constrain and disfavor the shallower $\alpha\lesssim 6$ values. Future broad-band monitoring will probe a larger portion of the parameter space of the KN fastest ejecta (\S\ref{SubSec:KNejecta}). 
} \label{fig:Xray}
\end{center}
\end{figure*}

For each observation ID 
we extracted a spectrum using \texttt{specextract} and a source region of 1.5\arcsec\,  centered at the location of the X-ray counterpart. We fitted each spectrum using an absorbed power-law model (\texttt{tbabs*ztbabs*pow}) within \texttt{XSPEC} (v12.9.1), adopting a Galactic neutral hydrogen column density NH$_{MW} = 0.0784 \times 10^{22}$ cm$^{-2}$ (\citealt{Kalberla2005}). We employed Cash statistics and performed a series of Markov Chain Monte Carlo (MCMC) simulations to properly constrain the spectral parameters and their uncertainties in the regime of low-count statistics as in \cite{margutti2017electromagnetic,Margutti+2018,Alexander+2018}. In no case did we find any    
statistical evidence for significant intrinsic absorption N$_{H,\rm{int}}$, and we list the derived $3\,\sigma$ upper limits in Table \ref{tab:wavdetecttab1}. We thus assume N$_{H,\rm{int}}=0\,\rm{cm^{-2}}$ in our subsequent modeling. The inferred best-fitting photon indices $\Gamma$, absorbed fluxes and (unabsorbed) luminosities are reported in Table \ref{tab:wavdetecttab1}. For observations acquired within a few days of each other, we also provide the results from a joint spectral fit and we plot the resulting light-curve in Fig. \ref{fig:Xray}. Finally we do not find statistical evidence for spectral evolution of the source over $\delta t \sim 2-743$ days. From a joint-fit of all the CXO data at $\delta t\ge9.2$ days we infer N$_{H,\rm{int}} < 0.69 \times 10^{22}$ cm$^{-2}$ and $\Gamma = 1.57{^{+0.12}_{-0.07}}$ (for N$_{H,\rm{int}}=0\,\rm{cm^{-2}}$), consistent with the spectral index inferred from broad-band radio to X-ray studies (e.g. \citealt{Margutti+2018,Alexander+2018,troja+2018,Fong19}).

\begin{deluxetable*}{cccccccccccc}
\tablecaption{\label{tab:wavdetecttab1} Results from our homogeneous spectral analysis of all the CXO observations of GW\,170817 between $2.3$ and $743$ days since merger. The reported \\photon indices, absorbed fluxes and (unabsorbed) luminosities are calculated for N$_{H,\rm{int}}=0\,\rm{cm^{-2}}$. At $\delta t>400$ days the photon index $\Gamma$ is not well constrained\\ and we adopt $\Gamma=1.57$ for the spectral calibration.  The reported significance is for a blind (targeted) detection for $\delta t<360$ days ($\delta t>360$ days).}
\tabletypesize{\footnotesize}
\tablecolumns{12} 
\tablewidth{\textwidth}
\tablehead{ 
\colhead{} & \colhead{Time since} & \colhead{Significance} & \colhead{Exposure} & \colhead{Net count rate} & \colhead{N$_{H,\rm{int}}$} & \multicolumn{2}{c}{Photon Index}  &\multicolumn{2}{c}{Absorbed Flux}  & \multicolumn{2}{c}{Luminosity} \\
\colhead{ObsID} & \colhead{merger}  & \colhead{$\sigma$}  & \colhead{(ks)}  & \colhead{(0.5-8 keV)}  &\colhead{3-$\sigma$ upper limit} & \multicolumn{2}{c}{$\Gamma$}& \multicolumn{2}{c}{(0.3 - 10 keV)}  & \multicolumn{2}{c}{(0.3-10 keV)}  \\
\colhead{} & \colhead{(days)} & \colhead{} & \colhead{} & \colhead{(10$^{-4}$ cts/s)} & \colhead{10$^{22}$ cm$^{-2}$} & \multicolumn{2}{c}{}  &\multicolumn{2}{c}{10$^{-15} $erg cm$^{-2}$s$^{-1}$} & \multicolumn{2}{c}{10$^{38}$ erg s$^{-1}$} }
\startdata
18955 & 2.33& --& 24.6&$<1.2$ &  --& \multicolumn{2}{c}{2 (assumed)} &\multicolumn{2}{c}{$<1.4$}  &\multicolumn{2}{c}{ $<3.4$}\\[5pt]
19294 & 9.20 & 5.8 &  49.4 & 2.9 $\pm$ 0.8 & $<$ 17.6 &\multicolumn{2}{c}{0.91$_{-0.39}^{+0.94}$}   & \multicolumn{2}{c}{6.85$^{+3.20}_{-3.04}$}   &\multicolumn{2}{c}{13.50$^{+6.31}_{-6.00}$}\\[5pt] 
20728 & 15.38 & 7.2 &  46.7&  3.8 $\pm$ 0.9 & \rdelim\}{2}{*}[$<$ 6.1] & 1.55$^{+0.76}_{-0.33}$ &\rdelim\}{2}{*}[2.22$_{-0.35}^{+0.76}$] & 5.99$^{+2.54}_{-1.02}$ &\rdelim\}{2}{*}[4.32$^{+1.15}_{-0.56}$] & 12.53$^{+5.31}_{-2.14}$ & \rdelim\}{2}{*}[10.35$^{+2.76}_{-1.35}$]\\[3pt]
18988 & 15.94 & 5.3 & 46.7 &  3.0 $\pm$ 0.8 & & 3.54$^{+1.49}_{-0.44}$ & & 5.55$^{+2.11}_{-1.95}$ & & 19.37$^{+7.37}_{-6.80}$ & \\[7pt]
20860 & 107.97 & 33.4  & 74.1 &  14.7 $\pm$ 1.4 & \rdelim\}{2}{*}[$<$ 1.7] & 1.48$^{+0.22}_{-0.14}$& \rdelim\}{2}{*}[1.52$_{-0.12}^{+0.17}$] & 25.19$^{+1.63}_{-2.04}$ & \rdelim\}{2}{*}[24.24$^{+2.09}_{-2.87}$] &52.20$^{+3.38}_{-4.22}$ & \rdelim\}{2}{*}[50.48$^{+4.35}_{-5.98}$]\\[3pt]
20861 & 111.06 & 14.9 & 24.7 & 14.1 $\pm$ 2.4 & & 1.67$^{+0.44}_{-0.28}$&& 21.07$^{+6.38}_{-3.70}$ & & 44.87$^{+13.60}_{-7.87}$ &\\[7pt]
20936 & 153.55 & 22.5  &  31.8& 18.6 $\pm$ 2.5 & \rdelim\}{6}{*}[$<$ 1.2] & 1.35$^{+0.31}_{-0.19}$& \rdelim\}{6}{*}[1.58$_{-0.15}^{+0.26}$] & 32.37$^{+13.89}_{-8.83}$ & \rdelim\}{6}{*}[24.20$^{+3.60}_{-1.71}$] & 66.00$^{+28.33}_{-18.00}$ & \rdelim\}{6}{*}[50.84$^{+7.57}_{-3.59}$]\\[3pt]
20938 & 157.12& 13.5  & 15.9 & 18.5 $\pm$ 3.5 & & 1.75$^{+0.46}_{-0.24}$ & & 26.46$^{+4.11}_{-6.17}$ & & 57.14$^{+8.87}_{-13.33}$ &\\ [3pt]
20937 & 158.92 & 12.6 & 20.8 & 13.6 $\pm$ 2.6 & &1.90$^{+0.57}_{-0.26}$ & & 20.39$^{+6.76}_{-2.00}$ & & 45.34$^{+15.04}_{-4.44}$ &\\[3pt]
20939 & 159.93 & 10.6 & 22.2 &  10.8 $\pm$ 2.3 & & 1.93$^{+0.61}_{-0.40}$& & 14.37$^{+1.42}_{-1.75}$ & & 32.16$^{+3.19}_{-3.91}$ &\\[3pt]
20945 & 163.73 & 7.4 & 14.2 & 11.5 $\pm$ 2.9 & & 1.61$^{+1.03}_{-0.42}$ & & 21.06$^{+6.11}_{-7.56}$ & &44.40$^{+12.88}_{-15.94}$ &\\[7pt]
21080 & 259.20 & 13.8  & 50.8 &  7.8 $\pm$ 1.3 & \rdelim\}{2}{*}[$<$ 3.8] & 1.62$^{+0.44}_{-0.27}$& \rdelim\}{2}{*}[1.57$_{-0.13}^{+0.29}$] & 11.40$^{+1.86}_{-1.32}$ & \rdelim\}{2}{*}[12.21$^{+2.88}_{-1.48}$] & 24.11$^{+3.93}_{-2.80}$ & \rdelim\}{2}{*}[25.61$^{+6.04}_{-3.10}$]\\[3pt]
21090 & 260.78 & 14.8 & 46.0 & 8.3 $\pm$ 1.4 & & 1.52$^{+0.35}_{-0.24}$ & & 13.06$^{+4.58}_{-1.52}$ & &27.22$^{+9.53}_{-3.18}$&\\ [5pt]
21371 & 358.61 & 11.1  & 67.2 & 5.0 $\pm$ 0.9 & $<$ 3.9 & \multicolumn{2}{c}{1.69$^{+0.49}_{-0.34}$} &  \multicolumn{2}{c}{7.75$^{+2.70}_{-0.73}$} & \multicolumn{2}{c}{16.58$^{+5.77}_{-1.56}$}\\[5pt] 
21322 & 580.99  & 2.3 &35.6 & 1.5 $\pm$  0.7 & \rdelim\}{3}{*}[$<$ 3.05] & $0.95^{+1.73}_{-1.67}$ &\rdelim\}{3}{*}[1.28$_{-0.15}^{+1.04}$] & 5.43${^{+2.54}_{-2.20}}$ &\rdelim\}{3}{*}[3.25${^{+0.85}_{-1.03}}$] & 11.11${^{+5.19}_{-4.49}}$ & \rdelim\}{3}{*}[7.07${^{+1.86}_{-2.24}}$]\\[3pt]
22157 & 581.94 & 2.7 &38.2  & 1.6  $\pm$  0.7 & & $1.38^{+1.09}_{-1.05}$ & & 2.75${^{+2.20}_{-1.36}}$ & &5.82${^{+2.43}_{-2.57}}$&\\[3pt]
22158 & 583.60  & 2.0 & 24.9 & 1.5$\pm$0.8 & & $1.59^{+2.82}_{-2.57}$& & 5.75${^{+3.78}_{-3.01}}$ & &12.53${^{+8.24}_{-6.57}}$&\\[5pt]
21372 & 740.31 & 2.2 &40.0 & <1.3  & \rdelim\}{3}{*}[$<$ 11.4] & -- &\rdelim\}{3}{*}[$1.23^{+1.05}_{-1.03}$] & -- & \rdelim\}{3}{*}[2.21${^{+0.85}_{-0.79}}$] & -- & \rdelim\}{3}{*}[4.82${^{+1.86}_{-1.71}}$]\\[3pt]
22736 & 742.26 & 3.0  &33.6 & $1.0 \pm 0.4$ & & $2.61^{+2.66}_{-2.01}$& & 2.45${^{+1.39}_{-1.62}}$ & & 6.77${^{+3.85}_{-4.49}}$ & \\[3pt]
22737 & 743.13 &4.6 & 25.2 & $2.2 \pm 0.9$ & & $1.21^{+1.46}_{-1.45}$ & &7.22${^{+3.25}_{-2.95}}$  & & 15.03${^{+6.78}_{-6.15}}$  & \\[3pt]
\hline
Joint Fit & -- & --& -- & -- & $<$ 0.69 & 1.57 $^{+0.12}_{-0.07}$ & & --  & &  -- \\
\enddata
\end{deluxetable*}

\subsection{Spatially Resolved Spectral Analysis of the Host Galaxy Diffuse X-ray Emission}
\label{SubSec:newmethod}

\begin{figure}[!t]
\begin{center}
\hspace*{-0.1in}
\scalebox{1.}
{\includegraphics[width=0.45\textwidth]{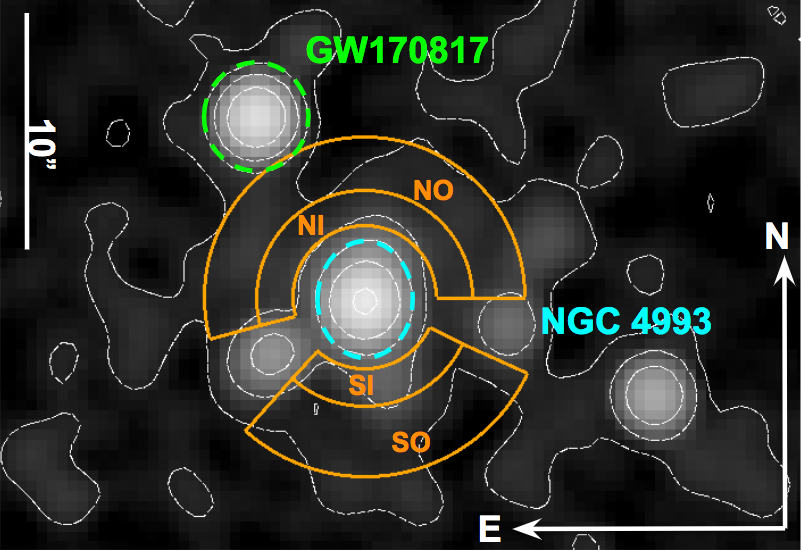}}
\caption{Combined X-ray image with CXO observations from $\delta$t$\sim$ 2 days - 743 days post-merger in 0.5--8 keV energy range, with contour levels in white. The four sectors in orange (NI: North Inner, NO: North Outer, SI: South Inner, SO: South Outer) mark the regions of spectra extraction for the spatially resolved X-ray spectral analysis of \S\ref{SubSec:newmethod}. The regions are defined so as to: (i) exclude emission from the core of the host galaxy, (ii) avoid contamination from the neighboring point-sources, and (iii) have comparable number of background subtracted counts.} 
\label{Fig:diffuse}
\end{center}
\end{figure}
The host galaxy of GW\,170817 (NGC4993) shows evidence for diffuse X-ray emission from a hot interstellar medium (ISM), in addition to harboring a weak active galactic nucleus (AGN, e.g. \citealt{blanchard2017electromagnetic})
and point sources of X-ray emission (Fig. \ref{Fig:diffuse}). In this section we describe the results from a spatially resolved X-ray spectral analysis of NGC\,4993, with the goal to constrain the physical properties of the plasma responsible for the diffuse emission component, (i.e. plasma temperature $T$ and particle density $n$), taking advantage of the very deep merged CXO observation. 
 
We followed the method  developed by \cite{Paggi14} to constrain the physical properties of the hot ISM of the  elliptical galaxy NGC\,4649. As a first step, we merged all the observations (with a total exposure time of $\sim 731$ ks) into a single event file using the \texttt{merge\_obs} task within \texttt{CIAO}. \texttt{merge\_obs} integrates two separate tasks: \texttt{reproj\_obs}, which re-projects individual event files to a common astrometric solution, and \texttt{flux\_obs}, which then merges the re-projected files into a single exposure-corrected event file. Other products from \texttt{merge\_obs} include re-projected images, exposure maps and exposure-corrected images in a given energy band.
We then combined the point spread function (PSF) maps of individual observations into a single exposure-map weighted PSF file with \texttt{dmimgcalc}. Finally, we used the exposure-map weighted PSF file from the previous step, the merged re-projected 0.5-8 keV event file, and the exposure map created by \texttt{merge\_obs} as input to  \texttt{wavdetect}. Our goal was to detect faint point sources that would elude searches in individual exposures. We used a false-alert probability threshold of $4\times 10^{-6}$ and a set of different wavelet scales (i.e. 1, 2, 8 and 16). 
 Visual inspection reveals that this method reliably identifies all the sources of point-like X-ray emission in the merged image. The end product of this process is a list of detected point sources and corresponding point-source regions. 
 
 We defined four regions for the extraction of the spectra of the diffuse X-ray emission as in Fig. \ref{Fig:diffuse}. The inner sectors  (NI and SI in Fig. \ref{Fig:diffuse})  have an internal radius $r_{i,1}=3.5\arcsec$ (to exclude the emission from the host galaxy core, which is dominated by the AGN), an external radius $r_{e,1}=5.25\arcsec$, and angular extents defined to avoid the point sources identified above. The north-outer sector (NO in Fig.  \ref{Fig:diffuse}) has an inner radius $r_{i,2}=r_{e,1}$ and extends to $r_{e,2}=7.8\arcsec$ (near GW\,170817). The SO sector extends from $r_{e,1}$ to $r_{e,2}=8.7\arcsec$. These regions are defined so as to contain a number of photons that lead to $\gtrsim3 \sigma$ evidence for emission in excess to the expected background counts, which corresponds to $N\sim 20-50$ 
background-subtracted counts in the different regions.

For each observation ID, we extract four spectra with \texttt{specextract} (one for each of the sectors of Fig. \ref{Fig:diffuse}) with the background spectra extracted from the nearby `blank-sky' field, generating spectral response files that are weighted by the count distribution within the aperture, as appropriate for extended sources. Finally, for each sector, we combined the spectral files obtained in the previous step using \texttt{combine\_spectra}.

We modeled the emission from hot plasma in NGC\,4993 with the \texttt{apec} model within \emph{Xspec}. Due to projection effects, each 2-D sector in Fig. \ref{Fig:diffuse} collects part of the radiation from 3-D shells at larger radii. We accounted for these projection effects using the \texttt{projct} mixing model within \emph{Xspec}, that is designed to perform a 3-D to 2-D projection of prolate ellipsoidal shells onto elliptical annuli (and respective sectors).
We further adopted the solar abundances  
from \cite{Asplund2009} and fixed the metal abundance parameter of the \texttt{apec} model to three different values of 0.5\,Z$_\odot$, Z$_\odot$, 2\,Z$_\odot$, (where Z$_\odot$ is the solar metallicity).
The galactic absorption column density was frozen to NH$_{MW} = 0.0784 \times 10^{22}$ cm$^{-2}$ (\citealt{Kalberla2005}) for all the spectral fits. The fit was initially performed for the outermost sectors (NO and SO in  Fig. \ref{Fig:diffuse}) independently, and the best-fitting parameters are reported in Table \ref{tab:deprojectedannuli}. The fit of the inner sectors were then performed jointly with their respective outer sectors, with the spectral parameters of the outer sectors frozen to the best-fitting parameters obtained in the previous step. 
All the resulting best-fitting de-projected model parameters (i.e. plasma temperature and emission measure EM) for each sector are presented in Table \ref{tab:deprojectedannuli} for three  metallicity values.

\subsubsection{Inferred ionized matter density at the location of GW\,170817}

The best-fitting EM value of the \texttt{apec} diffuse 
emission model for the different shells provides a direct estimate of the host-galaxy density at that location. The EM is defined as:

\begin{equation}
\label{eq:eq1}
EM = \frac{10^{-14}}{4\pi D_A^2}\int n_e n_H dV \approx \frac{10^{-14}}{4\pi D_A^2} \left(\frac{\rho}{m_p}\right)^2\frac{X(1+X)}{2}V_{olume} 
\end{equation}
where  
$D_A$ is the angular distance to the host galaxy (in cm); $n_e$ and $n_H$ are the number density of electrons and hydrogen atoms (in cm$^{-3}$), respectively, and $n_e \sim \frac{\rho}{2m_p}(1+X)$. $\rho$ is the matter density, $X$ is the fraction of hydrogen by mass, and $m_p$ is the proton mass. The particle densities inferred from the de-projected \texttt{apec} spectral fits are reported in Table \ref{tab:deprojectedannuli}. Of particular interest are the density values inferred for the outer sector NO. We find $n\sim (5.1-9.6)\times10^{-3}\,\rm{cm^{-3}}$, depending on the assumed gas metallicity. GW\,170817 is located at larger radius (Fig. \ref{Fig:diffuse}), where the gas density is likely to be lower. Additionally, unresolved point sources might contribute some of the detected emission. We thus consider  
$n\le9.6 \times 10^{-3} \,\rm{cm^{-3}}$ as an  upper limit on the density of ionized matter in the merger environment. Our density constraint analysis is not sensitive to the presence of small-scale density variations, for instance, the presence of over-densities at the edge of a bow-shock cavity formed if the merger progenitor hosted a pulsar (e.g. \citealt{Ramirez+2019}). Our analysis complements previous inferences of neutral hydrogen particle density  $n_{HI}<0.04\,\rm{cm^{-3}}$ derived from radio observations by \cite{hallinan2017counterpart}, and it is consistent with the lower-limit on the circum-merger density  $n>2\times 10^{-5}\,\rm{cm^{-3}}$ derived by \cite{Mooley1+2018}.

\begin{table}[!htb]
\setlength\tabcolsep{0pt}
\caption{Best-fitting de-projected emission measure (EM) and temperature $T$ derived from a bremsstrahlung spectral fit of the emission from the concentric annular regions of Fig. \ref{Fig:diffuse}, and derived particle density $n$. } \label{tab:deprojectedannuli}
\centering

\smallskip
\begin{tabular*}{\columnwidth}{@{\extracolsep{\fill}}rccccr}
\toprule
Shell & {EM\footnote{$EM = \frac{10^{-14}}{4\pi{D_A}^2}\int n_e n_H dV$, where $D_A$ is the angular diameter distance to the source (cm), $n_e$ and $n_H$ are the electron and H densities (cm$^{-3}$), respectively.}}& {Temperature ($T$)} & {C-stat/dof} & {Density ($n$)} \\
& ($\times 10^{-7}$ cm$^{-5}$) & (keV) & & ($\times 10^{-3}$ cm$^{-3}$)\\
\hline
\multicolumn{5}{c}{Z = 0.5 \(Z_\odot\)} \\
\hline
NO & 4.63$^{+1.33}_{-0.79}$& 0.68$^{+0.07}_{-0.11}$& 386/510 & 9.60$^{+1.38}_{-0.82}$ \\[2pt]
NI & 12.41$^{+3.59}_{-3.27}$& 2.03$^{+1.49}_{-0.62}$& 679/1022 & 28.29$^{+4.09}_{-3.73}$\\[2pt]
SO & 2.71$^{+1.14}_{-1.07}$& 1.83$^{+3.62}_{-0.67}$& 296/510 & 7.68$^{+1.62}_{-1.51}$\\[2pt]
SI & 30.74$^{+6.37}_{-5.55}$&  4.28$^{+8.59}_{-1.84}$& 539/1022 & 58.06$^{+6.02}_{-5.24}$\\[2pt]
\hline
\multicolumn{5}{c}{Z = \(Z_\odot\)} \\
\hline
NO & 2.57$^{+0.69}_{-0.45}$ & 0.68$^{+0.07}_{-0.11}$ & 385/510 & 7.15$^{+0.97}_{-0.69}$ \\[2pt]
NI & 9.72$^{+3.08}_{-3.05}$& 2.18$^{+1.81}_{-0.62}$ & 679/1022 & 25.04$^{+3.96}_{-3.93}$\\[2pt]
SO & 2.36$^{+1.07}_{-1.20}$ & 2.67$^{+4.65}_{-1.39}$& 296/510& 7.16$^{+1.63}_{-1.82}$\\[2pt]
SI & 11.59$^{+2.74}_{-2.41}$& 0.74$^{+0.25}_{-0.16}$& 547/1022 & 35.66$^{+4.21}_{-3.70}$\\[2pt]
\hline
\multicolumn{5}{c}{Z = 2 \(Z_\odot\)} \\
\hline
NO & 1.36$^{+0.36}_{-0.24}$ & 0.68$^{+0.07}_{-0.11}$& 385/510 & 5.19$^{+0.69}_{-0.46}$\\[2pt]
NI & 7.03$^{+2.79}_{-2.66}$ & 2.43$^{+2.09}_{-0.81}$ & 679/1022  & 21.28$^{+4.24}_{-4.02}$\\[2pt]
SO & 2.15$^{+0.84}_{-1.09}$& 4.29$^{+3.68}_{-2.60}$& 296/510 & 7.37$^{+1.56}_{-1.95}$\\[2pt]
SI & 3.26$^{+0.79}_{-0.64}$ & 0.64$^{+0.12}_{-0.11}$ & 490/1022 & 18.90$^{+2.29}_{-1.84}$\\[2pt]
\bottomrule
\end{tabular*}
\end{table}

\section{Radio Data Analysis}
\label{sec:radio}
We observed GW170817 with the Karl G. Jansky Very Large Array (VLA) on 2019 January 21 beginning at 12:32:10 UT ($\delta t \sim 521$ days post merger), 2019 January 25 at 10:52:45 UT ($\delta t \sim 525$ days), and 2019 March 29 at 05:00:15 UT ($\delta t \sim 588$ days). The January observations lasted 2 hours each and were taken in C configuration, while the March observation lasted 4 hours and was taken in B configuration. All observations were taken at a mean frequency of 6 GHz with an observing bandwidth of 4 GHz. The data were calibrated and imaged with standard CASA routines \citep{casa}, using 3C286 as the flux calibrator and J1258-2219 as the phase calibrator.
 
We do not detect GW170817 in any of the observations individually or in a combined image made from the two January observations. We therefore combine all three datasets using the CASA task {\tt concat} and  produce a single image with improved signal-to-noise. We recover a faint source at the location of GW170817 in the final joint image. We fit the emission with a point source model using the {\tt imtool} package within {\tt pwkit} \citep{Williams+17} and obtain a final flux density of $5.9 \pm 1.9$ $\mu$Jy. This is consistent with expectations for an off-axis structured relativistic jet, Fig. \ref{fig:Xray} \citep{xie2018,Alexander+2018,WuMacfadyen2018}. 

A final epoch of radio observation was acquired at $\delta t=724.3-743.2$ days since merger, and consisted of two observations, the first beginning on 2019 August 11 at 19:36:09 UT (3 hours, A configuration)
and the second beginning on 2019 August 30 at 18:29:44 UT (3 hours, A configuration).
For both observations the mean frequency is 6 GHz and the bandwidth is 4 GHz. Following the same data reduction and calibration procedure as above we do not find evidence of radio emission at the location of GW\,170817 in the individual observations or in a combined image. 
We also imaged the output of the observatory-provided NRAO pipeline calibrated data and obtained similar results. We infer  $F_{\nu}<8.4\,\rm{\mu Jy}$  at $3\,\sigma$ c.l. from the combined dataset. We show the complete 6 GHz radio lightcurve of GW\,170817 in Fig. \ref{fig:Xray}.

The radio to X-ray SED at $\delta t \sim 582$ days is well modeled by a simple power-law with $F_{\nu}\propto \nu^{-\beta}$ and $\beta=0.55\pm0.02$ consistent with \cite{Fong19} and the inferred broad-band spectrum at earlier times (e.g. \citealt{Alexander+2018,Davanzo2018,Margutti+2018,Mooley3+2018, Dobie+2018,troja+2018,Troja2019}). We further infer a 3$\sigma$ lower limit on the synchrotron cooling break frequency $\nu_c>0.16$ keV 
at $\delta t \sim 582$ days. Based on data presented in this section and \S\ref{Sec:CXOdata} we conclude that there is no evidence for spectral evolution of the non-thermal emission of GW\,170817 at any time of our monitoring, from $\delta t \sim 10$ days until $\sim740$ days since merger.

\section{Updated modeling of the broad-band jet afterglow emission}
\label{SubSec:afterglow}
We use \texttt{JetFit}, the synthetic light-curve fitting tool based on the two-parameter boosted fireball model developed by \cite{Duffell+2013} and \cite{WuMacfadyen2018}, to fit the broad-band non-thermal emission from GW\,170817 up to $\sim2$ yrs since merger. \texttt{JetFit} can naturally accommodate a wide range of outflow structures ranging from mildly relativistic quasi-spherical outflows to ultra-relativistic structured jets \citep{WuMacfadyen2018,WuMacFadyen2019}.
 Specifically, our data set consists of the X-ray observations from Table \ref{tab:wavdetecttab1}, $\sim$3\,GHz and $\sim$6\,GHz VLA radio observations collected from \cite{Alexander+2017}, \cite{hallinan2017counterpart}, \cite{Mooley3+2018}, \cite{Margutti+2018}, \cite{Alexander+2018}, \cite{Dobie+2018}, \cite{Mooley1+2018}, as well as our latest radio observations presented in Sec.\ref{sec:radio}. 
 
\begin{figure*}[!t]
\begin{center}
\hspace*{-0.1in}
\scalebox{1.}
{\includegraphics[width=\textwidth]{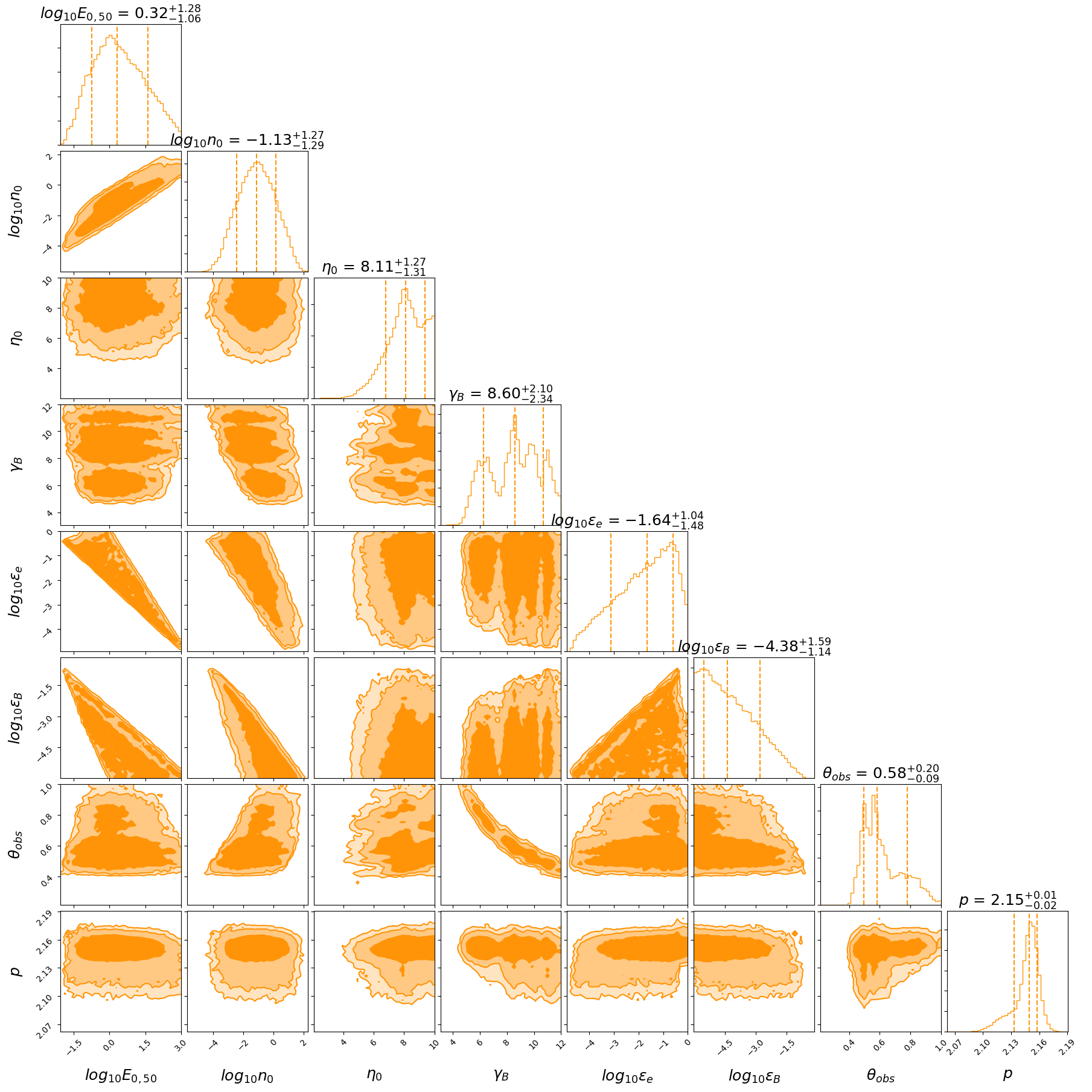}}
\caption{Corner plot showing the one and two dimensional projection of the posterior probability distribution of the jetted-outflow model parameters. Vertical dashed lines in the one-dimensional projections of the posterior distribution mark the 16\%, 50\% and  84\% percentiles of the marginalized distributions, (i.e. the median value and the $1\,\sigma$ range).  The contours are drawn at 68\%,95\%, and 99\% confidence levels.}
\label{fig:contourcorner}
\end{center}
\end{figure*}

Within \texttt{JetFit} the synthetic light curves are generated using four hydrodynamical paramaters: explosion energy $E_0$ (one side), ambient density $n$, asymptotic Lorentz factor $\eta_0$, and boost Lorentz factor $\gamma_B$; four radiative parameters: spectral index $p$ of the electron distribution $N_e(\gamma_e)\propto \gamma_e^{-p}$, the electron energy fraction $\epsilon_e$, the magnetic energy fraction $\epsilon_B$ and the fraction of electrons accelerated in a power-law distribution by the shock $\xi_N$; and three observational parameters: redshift $z$, luminosity distance $d_L$ and the observer angle $\theta_{obs}$ with respect to the launch direction of the fireball. Model parameters inferred from the synchrotron emission intrinsically suffer from a level of degeneracy due to the unknown $\xi_N$ value (e.g. \citealt{Eichler05}). We thus assume $\xi_N=1$ as common practice in the Gamma-Ray Burst (GRB) literature to allow a direct comparison to parameters inferred for short GRBs. 
We set the bounds on priors for the remaining eight parameters similar to those of \cite{WuMacfadyen2018} as reported in Table \ref{tab:medianparams}.
We perform MCMC fitting using 100 walkers and $10^4$ burn-in iterations. 
Sampling is performed on  $10^4$ additional iterations. The posterior distribution of the model parameters is generated with the \texttt{emcee} package \citep{Foreman-Mackey13}.
The one-dimensional and two-dimensional projections of the posterior distribution that result from our fits are shown in Fig. \ref{fig:contourcorner}, and the best-fitting model is shown in Fig. \ref{fig:Xray}. 
The median values of the fitting parameters are reported in Table \ref{tab:medianparams} with $1\sigma$ uncertainties computed as the 16th and 84th percentiles of the one-dimensional projection of the posterior distribution. These model parameters are consistent with those inferred by \cite{WuMacfadyen2018} using data at $\delta t<300$ days. Since the new radio and X-ray observations that we present here are consistent with the extrapolation of the model by \cite{WuMacfadyen2018} at later times, this result is not surprising.

\begin{figure*}[!t]
\begin{center}
\hspace*{-0.1in}
\scalebox{1.}
{\includegraphics[width=\textwidth]{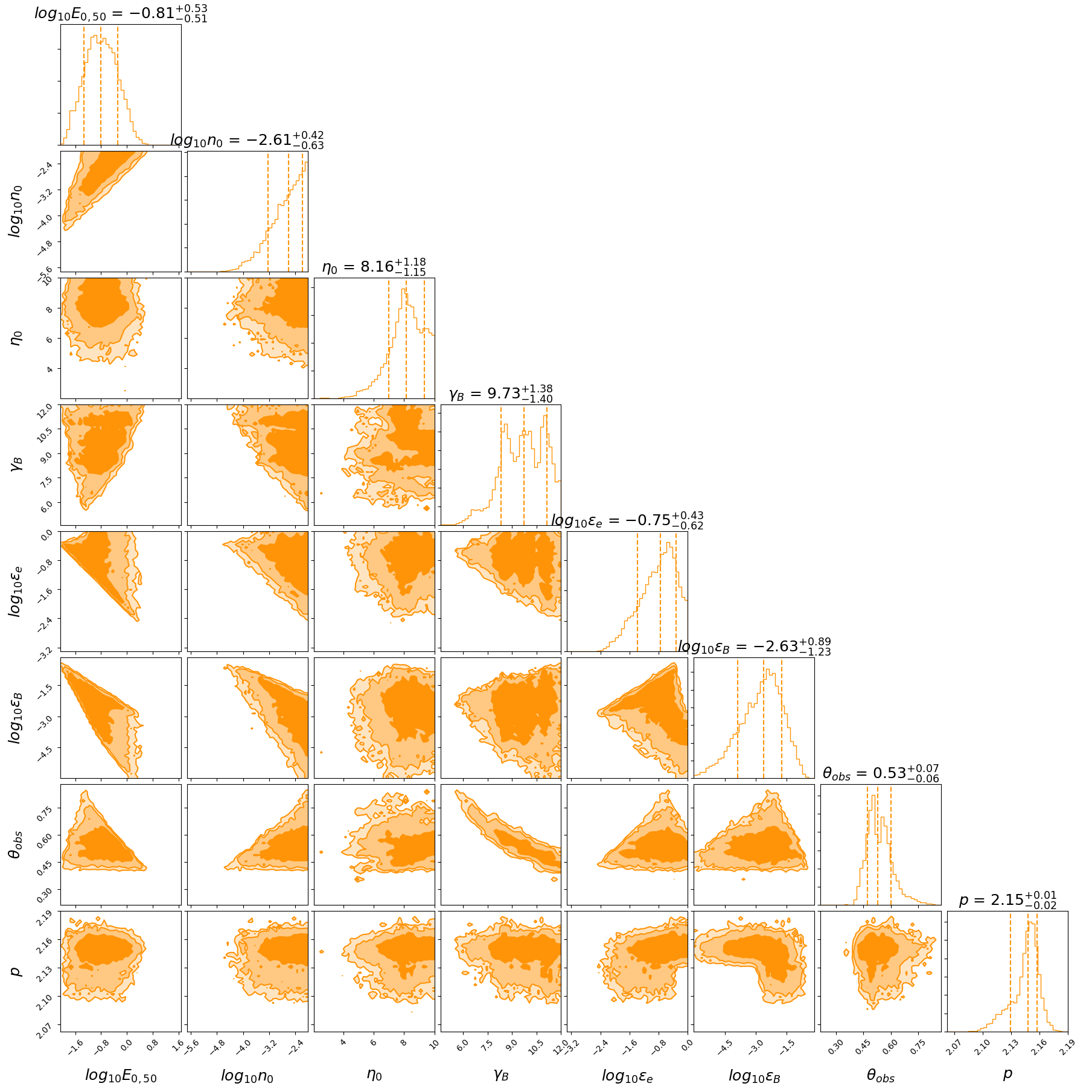}}
\caption{Same as Fig. \ref{fig:contourcorner} with the prior  $n\le9.6 \times 10^{-3}\,\rm{cm^{-3}}$ as found in \S\ref{SubSec:newmethod}.} 
\label{fig:contourdensUL}
\end{center}
\end{figure*}

The wide distributions of $E_0$ and $n$ (and $\epsilon_e$, and $\epsilon_B$) in Fig.\ref{fig:contourcorner} 
indicates a high level of degeneracy between the model parameters. As a refinement of our modeling, we enforce the upper limit on the ambient density of GW\,170817 derived in \S\ref{SubSec:newmethod}. From the posterior distribution derived above using \texttt{JetFit}, we reject all the samples with $n > 9.6 \times 10^{-3}\,\rm{cm^{-3}}$, and plot the revised distribution of parameters, as shown in Fig. \ref{fig:contourdensUL}, and the best-fitting model is shown in Fig\ref{fig:Xray}. The median values of the revised parameter distributions are reported in Table \ref{tab:medianparams}.
Taking the upper bound on the environment density into consideration when modeling the afterglow emission produces tighter constraints on the model parameters. 

We conclude that the broad-band non-thermal emission from GW\,170817  at $\sim 2$yr since merger (Fig. \ref{fig:Xray}) is still well described by an off-axis jetted-outflow model with angular structure. The outflow carries an explosion energy $E_0 \sim  1.5\times10^{49}\,\rm{erg}$  
(corresponding to an isotropic equivalent energy $E_{iso} \sim 2 \times 10^{52}\, \rm{erg} $), with a jet opening angle $\theta_0 \sim 6^{\circ}$, and characteristic Lorentz factor $\Gamma_j\sim 160$\footnote{The characteristic Lorentz factor of the outflow, $\Gamma_j$, mentioned here is different from $\Gamma$ that we used earlier to denote the photon index of the X-ray spectra in \S\ref{Sec:CXOdata}. When mentioned in reference to the kilonova, $\Gamma$ represents the Lorentz factor of the KN ejecta (as mentioned in Sec. \S\ref{SubSec:KNejecta}.)}
, expanding in a low-density environment ($n_0 \sim 2.5 \times 10^{-3}\, \rm{cm^{-3}}$). The jet axis is located at $\theta_{obs} \sim 30^{\circ} $ with respect to our line of sight. Our inferences are broadly consistent with structured jet model parameters from broad-band modeling attempts that included data extending to $\delta t\sim 300$ days (e.g \citealt{Ghirlanda2019,Lazzati2018,Mooley3+2018,Troja2019,Adithan+2019, Lamb+2019}).  
We find no evidence of departure from a steep post-peak light-curve decay and we infer $F_{\nu}\propto t^{-1.95 \pm 0.15}$ at $\delta t > 200$ days, consistent with previous findings at earlier times (e.g. \citealt{Alexander+2018,Mooley3+2018,Troja2019}) and the expectations from emission dominated by a collimated relativistic outflow seen off-axis \citep{Lamb+2018}.

The outflow will eventually enter the non-relativistic phase at $t_{NR}\propto (E_{k,iso}/n)^{1/3}$ (e.g. \citealt{Piran2004}), when the amount of swept-up material will be comparable to the kinetic energy of the outflow. The non-relativistic transition  will lead to a flattening of the light-curve decay $F_{\nu}\propto t^{-\alpha}$ with $\alpha=-(15p-21)/10\sim 1.1$ for  $\nu_m<\nu<\nu_c$ and $\alpha=-(3p-4)/2\sim 1.2$ above $\nu_c$ (e.g. \citealt{Huang2003,Gao2013}). For the outflow and environment density parameters listed in Table \ref{tab:medianparams}, the non-relativistic transition is expected to occur at $t_{NR}\sim 3600^{+2100}_{-2000}$ days ($t_{NR}\sim 4700^{+1700}_{-1400}$ days for the model with the $n\le 9.6 \times 10^{-3}\,\rm{cm^{-3}}$ prior). Before that happens, the KN afterglow might start dominating the observed emission (\S\ref{SubSec:KNejecta}).

\begin{table*}[!htb]
\setlength\tabcolsep{0pt}
\caption{\texttt{JetFit} model parameters and inferred quantities.} \label{tab:medianparams}
\centering
\smallskip
\begin{tabular*}{0.75\textwidth}{@{\extracolsep{\fill}}l c c c}
\toprule
& Bounds for & \multicolumn{2}{c}{Median value of}\\
Parameter & Prior Distribution\footnote{The priors on the parameters are taken as uniform distribution with the given bounds} & \multicolumn{2}{c}{Posterior Distribution} \\
 & & w/o density constraint  & w/ density constraint  \\
 \hline
 $\log_{10}E_{0,50}$ (erg) & [-6, 3] &$0.32_{-1.06}^{+1.28}$ &  $-0.81_{-0.51}^{+0.53}$\\
 $\log_{10}n_{0}$ (cm$^{-3}$) & [-6, 3] & $-1.13_{-1.29}^{+1.27}$ & $-2.61_{-0.63}^{+0.42}$\\
 $\log_{10} \epsilon_e$ & [-6, 0] & $-1.64_{-1.48}^{+1.04}$ & $-0.75_{-0.62}^{+0.43}$\\
 $\log_{10} \epsilon_B$ & [-6, 0] & $-4.38_{-1.14}^{+1.59}$ & $-2.63_{-1.23}^{+0.89}$\\
 $\eta_0$ & [2, 10] & $8.11_{-1.31}^{+1.27}$ & $8.16_{-1.15}^{+1.18}$ \\
 $\gamma_B$ & [1, 12] & $8.60_{-2.34}^{+2.10}$& $9.73_{-1.40}^{+1.38}$\\
 $\theta_{obs}$ (rad)  & [0, 1] & $0.58^{+0.20}_{-0.09}$& $0.53^{+0.07}_{-0.06}$\\
 $p$ & [2, 2.5] & $2.15^{+0.01}_{-0.02}$ & $2.15^{+0.01}_{-0.02}$ \\
\hline
\multicolumn{4}{c}{Derived Quantities}\\
\hline
$\theta_{0}$\footnote{$ \theta_{0}\sim 1/\gamma_B$} (deg) & & $6.66_{-1.31}^{+2.48}$ & $5.89_{-0.73}^{+0.99}$\\
$\log_{10}E_{iso,50}$\footnote{$E_{iso}\sim 2E_0/1-\cos{(\theta_0/2}$)}(erg) & & $3.34_{-1.07}^{+1.33}$ & $2.33_{-0.55}^{+0.60}$\\
$\Gamma_j$\footnote{$\Gamma_j\sim2 \eta_0\gamma_B$} & & $139_{-44}^{+39}$ & $163_{-43}^{+23}$\\
\bottomrule
\end{tabular*}
\end{table*}

\section{Constraints on the properties of the fastest KN ejecta}
\label{SubSec:KNejecta}

\begin{figure*}
\begin{center}
\hspace*{-0.4in}
\scalebox{1.}
{\includegraphics[width=0.25\textwidth]{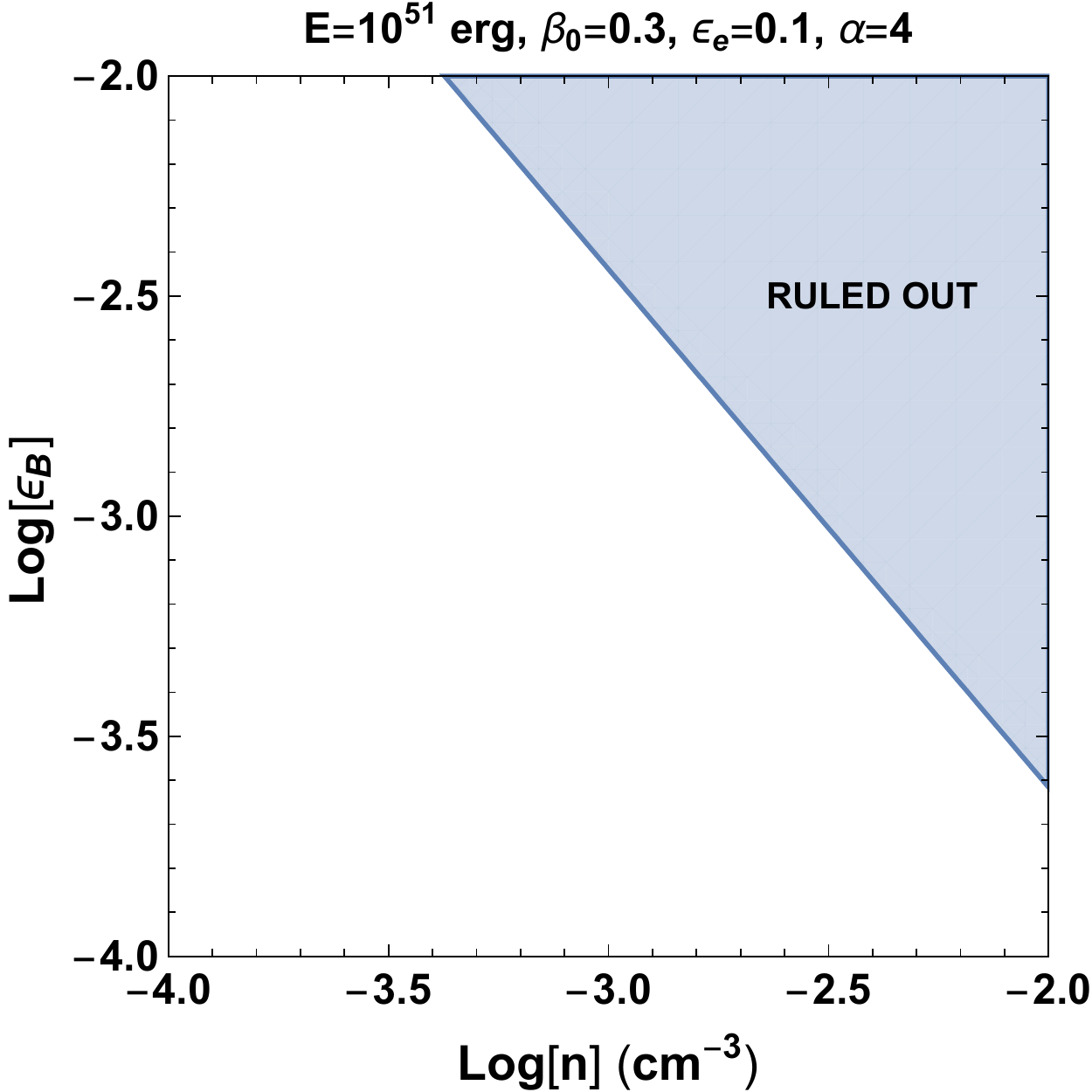}
\includegraphics[width=0.25\textwidth]{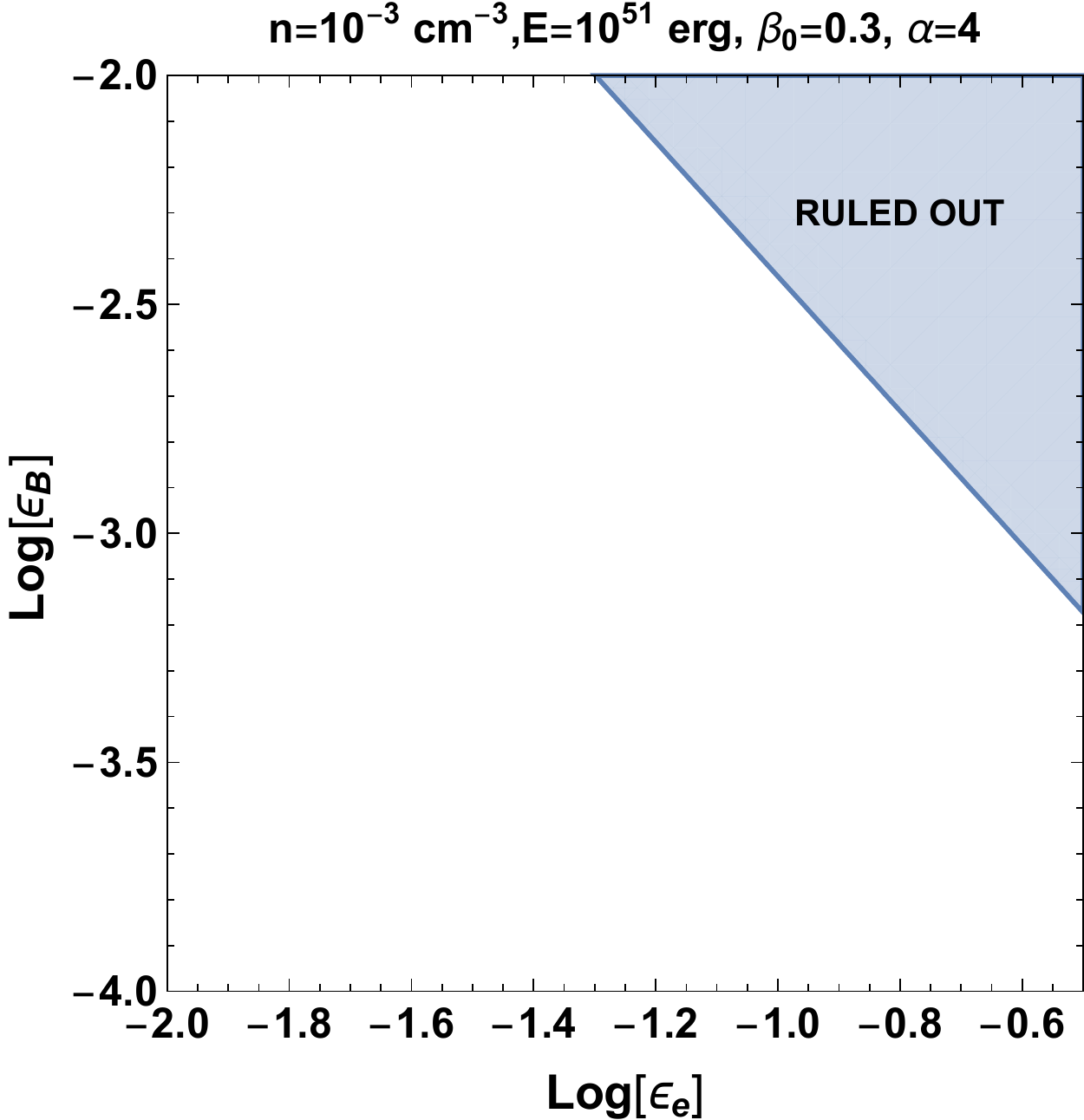}\includegraphics[width=0.25\textwidth]{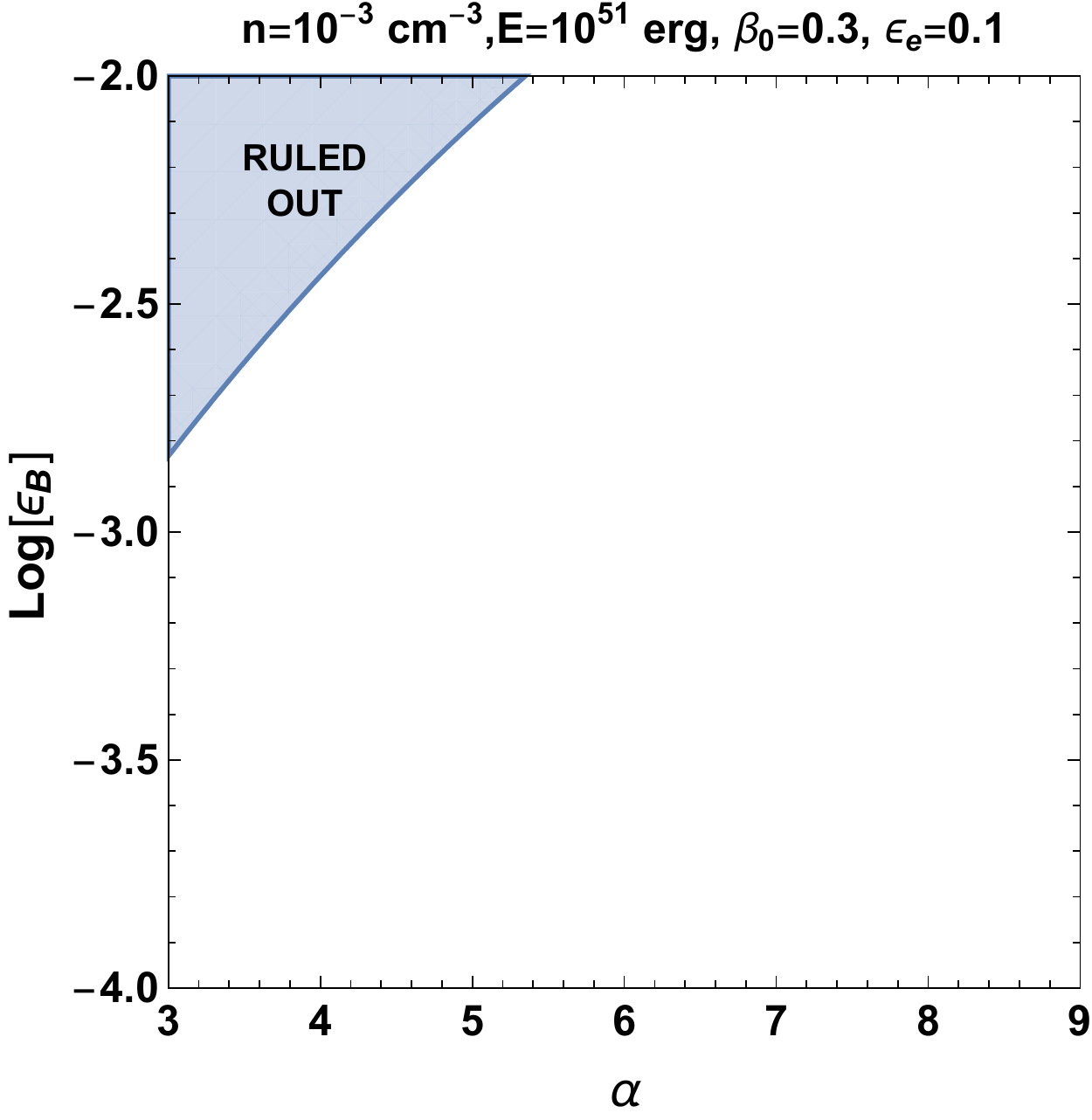}\includegraphics[width=0.25\textwidth]{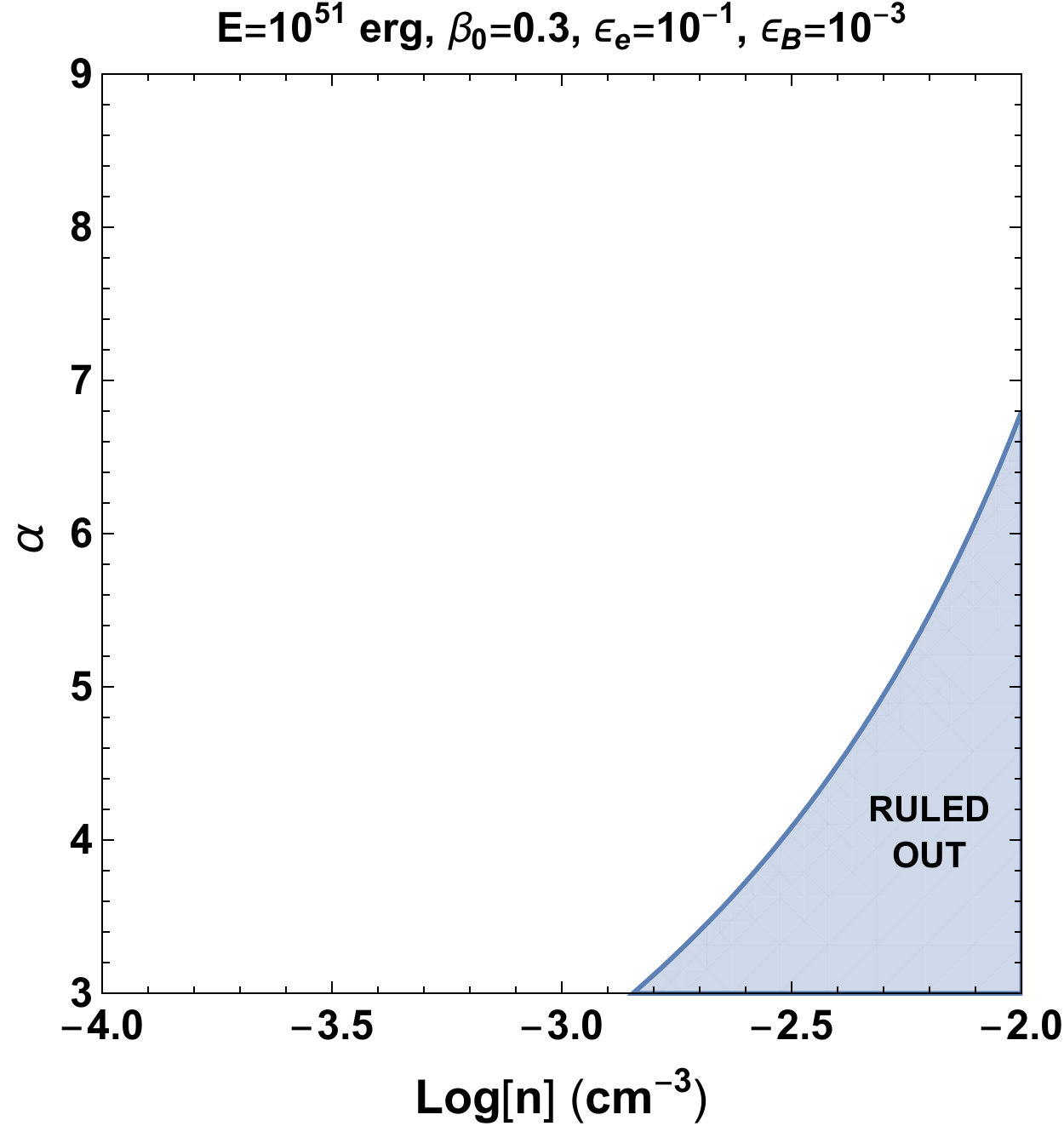}}
\caption{
Allowed (white) and ruled out (shaded) parameter space of the KN afterglow of GW\,170817 based on the fact that no re-brightening of the X-ray or radio emission was detected at $\sim2$ yrs after the merger (Fig. \ref{fig:Xray}). Two parameters are varied in each plot while the rest are kept fixed to values indicated in the plot title. The radio data set drives our conclusions and disfavors shallow stratification indexes $\alpha\le6$. 
Future observations will further constrain the parameter space of the KN afterglow.}
\label{fig:KNafterglow}
\end{center}
\end{figure*}
The deceleration of the KN ejecta into the ambient medium is another source of synchrotron radiation across the electromagnetic spectrum (i.e. the KN afterglow, e.g.  \citealt{nakarpiran2011}). In close analogy to stellar explosions, the bulk of the kinetic energy 
in KNe is carried by ``slowly'' moving material that powers the detected UV-optical-NIR KN  thermal emission, while the significantly lighter KN fastest ejecta rush ahead and shock the medium, accelerating electrons that cool via radiating synchrotron emission. By modeling the thermal UV-optical-NIR KN associated with GW\,170817, \cite{villar2017combined} constrained the bulk velocities and masses of the post-merger ejecta to $v\sim0.1-0.3c$ and total ejecta $M_{ej}\sim0.08\,\rm{M_{\odot}}$, carrying a kinetic energy in excess of $10^{51}\,\rm{erg}$ (see also \citealt{arcavi2017optical,cowperthwaite2017electromagnetic,drout2017light,evans2017swift,kasliwal2017illuminating,valenti2017discovery}). The KN thermal emission does not constrain the properties of the fastest KN ejecta at $\beta>0.3\,c$ and the velocity structure $E_k^{KN}(\Gamma \beta)$ of the KN outflow.  The kinetic energy profile $E_k^{KN}(\Gamma \beta)$ of the kilonova outflow 
carries direct information about the merger dynamics and, potentially, on the nature of the compact object remnant (e.g. \citealt{Hotokezaka18,Radice18,Radice18b,Fernandez19}). 

We parameterize the kinetic energy of the fastest KN ejecta as a power-law in specific momentum $\Gamma \beta$ with index $\alpha$: $E_k^{KN}(\Gamma \beta)\propto (\Gamma \beta)^{-\alpha}$ with a minimum outflow velocity $v_0$ motivated by thermal KM models. Following \cite{Kathirgamaraju19} we generated a set of broad-band KN afterglow light-curves with the typical parameters inferred for the afterglow of short gamma-ray bursts (GRBs): $v_0=0.3\,c$, total kinetic energy $\sim10^{51}\,\rm{erg}$, $p=2.2$, $\epsilon_e=0.1$, $\epsilon_B=[10^{-4}$--$10^{-2}]$, $n=[10^{-4}$--$10^{-2}]\,\rm{cm^{-3}}$ (\citealt{Fong+2015}), and with $\alpha=[3$--$9]$ (\citealt{Radice18,Radice18b}), which are shown in Fig.\ref{fig:Xray} along with the best-fitting off-axis structured jet models. We use the lack of evidence for emission from the KN afterglow to constrain the properties of the KN ejecta and its environment, as in \cite{Kathirgamaraju19}.  The results are displayed in Fig. \ref{fig:KNafterglow}, which shows that 
current radio observations disfavor shallow stratification indices $\alpha\le6$.
\footnote{We note that the KN afterglow and the jet afterglow do not necessarily share the same microphysical parameters $\epsilon_e$, $\epsilon_B$, and $p$ as the physical properties of the shocks launched by the two outflows are different.} Future observations at $\delta t\ge 1000$ days are more sensitive to the KN fastest ejecta tail and will probe a larger portion of the parameter space (Fig.\ref{fig:Xray}). 

\section{Summary and Conclusions}
\label{Sec:Conc}
We present deep X-ray and radio observations of GW\,170817 that extend to $\sim$2 yrs after the neutron-star merger, a homogeneous analysis of the entire X-ray data set, and a new method to independently constrain the density of the merger environment based on diffuse X-ray emission from hot plasma in the host galaxy. These observations offer a 
complete view of the evolution of the broad-band afterglow of an off-axis structured jet launched by the neutron star merger from its first detection at $\sim$10 days, peak at $\sim$160 days and steep decline until the present epoch, and place the first constraints on the properties of the kilonova (KN) afterglow. Our main results can be summarized as follows:
\begin{itemize}
\item Our analysis reveals no evidence for broad-band spectral evolution or temporal variability of the X-ray emission at any time. The radio-to-X-ray data are well described by a simple-power law spectrum $F_{\nu}\propto \nu^{-\beta}$ with $\beta = 0.575\pm0.007$. 
The highest statistical significance of short-term temporal X-ray variability is at the level of $2.5\,\sigma$.
\item From the analysis of diffuse X-ray emission from hot plasma in the host galaxy of GW\,170817 we infer a density limit on the NS merger environment $n\le 9.6 \times 10^{-3}\,\rm{cm^{-3}}$. We note however that our analysis does not capture small-scale variations in density.
\item  After $\sim$2 yrs of monitoring GW\,170817, we conclude that the non-thermal emission from the binary neutron-star merger has been dominated at all times by a jetted outflow with angular structure viewed off-axis (Fig.\ref{fig:Xray}). 
Modeling the afterglow emission without (with) the density constraint results in $\theta_{obs} = 33.2^{+11.5}_{-5.2}\, \rm{deg}$ ($\theta_{obs} = 30.4^{+4.0}_{-3.4}\, \rm{deg}$). The outflow carries E $ = 2.1_{-1.9}^{+38} \times 10^{50}\,\rm{erg}$ (E $ = 1.5^{+3.6}_{-1.1} \times 10^{49}\,\rm{erg}$) of energy and contains a core of collimated ultra-relativistic material (i.e. a jet) with inferred opening angle $\theta_0= 6.7_{-1.3}^{+2.5}\, \rm{deg}$  ($\theta_0= 5.9^{+1.0}_{-0.7}\, \rm{deg}$) and characteristic Lorentz factor $\Gamma_j= 139_{-44}^{+39}$ ($\Gamma_j= 163^{+23}_{-43}$). We infer an environment
density of $n = 7.3_{-6.9}^{+129} \times 10^{-2}\, \rm{cm^{-3}}$ ($n= 2.5^{+4.1}_{-1.9} \times 10^{-3}\, \rm{cm^{-3}}$).
We note that the values of opening angle, $\theta_0$ and spectral index, `p', of the electron distribution are the same in both scenarios.  
\item The lack of evidence of departure from the off-axis structured jet emission allows us to constrain the properties of the yet-to-be detected KN afterglow. We find that for fiducial values of the parameters of the KN ejecta kinetic energy distribution $E_k^{KN}(\Gamma\beta)\propto (\Gamma\beta)^{\alpha}$, current radio data disfavor shallow stratification indices $\alpha\le6$.
\end{itemize}

Future X-ray and radio observations of GW\,170817 have the potential to detect the very first electromagnetic signature of non-thermal emission from the deceleration of the fastest ejecta from a kilonova. Simulations show that the fastest KN ejecta is launched  by a shock when the merger remnant bounces back after merger (e.g. \citealt{Radice18b}). The detection of emission from a fast KN outflow  would (i) confirm that a high-mass neutron star was formed that was temporarily stable to collapse, ruling out prompt black hole formation; (ii) directly provide a constraint on the neutron star equation of state at higher densities than those probed by current LIGO/Virgo constraints on tidal deformability (as the process of ``bounce'' happens at higher densities and temperatures). 
\acknowledgements
R.M. thanks Kenta Hotokezaka for a question at a conference that inspired the search for a new method to constrain the density in the environment of GW170817 with deep X-ray observations.
The Margutti group at Northwestern acknowledges support provided by the National Aeronautics and Space Administration through grant HST-GO-15606.001-A, through Chandra Awards Number GO9-20058A, DD8-19101A and DDT-18096A issued by the Chandra X-ray Center, which is operated by the Smithsonian Astrophysical Observatory for and on behalf of the National Aeronautics Space Administration under contract NAS8-03060. Raffaella Margutti is a CIFAR Azrieli Global Scholar in the Gravity \& the Extreme Universe Program, 2019. R.M. acknowledges support by the National Science Foundation under Award No. AST-1909796. A.B. is supported by the Heising-Simons Foundation under grant  \#2018-0911 (PI: Margutti).
A.H. is supported by a Future Investigators in NASA Earth and Space Science and Technology (FINESST) award  19-ASTRO19-0158.
W.F. acknowledges support by the National Science Foundation under Award No.\ AST-1814782 and AST-1909358, and 
by the National Aeronautics and Space Administration through Chandra Award Number GO9-20058A issued by the Chandra X-ray Center, which is operated by the Smithsonian Astrophysical Observatory for and on behalf of the National Aeronautics Space Administration under contract NAS8-03060. 
K.D.A. acknowledges support provided by NASA through the NASA Hubble Fellowship grant HST-HF2-51403.001 awarded by the Space Telescope Science Institute, which is operated by the Association of Universities for Research in Astronomy, Inc., for NASA, under contract NAS5-26555.
R.C. acknowledges support from NASA Chandra grant GO9-20058B.
\facilities{CXO, VLA}
\software{HEASoft, CIAO, CASA, JetFit}
\newpage
\appendix
\renewcommand\thefigure{A\arabic{figure}}
\section{Blind search for Temporal Variability at $\delta t\sim 160$ \lowercase{days}}
\label{App:variability}

\begin{figure}[!t]
\begin{center}
\hspace*{-0.1in}
\scalebox{1.}
{\includegraphics[width=0.5\textwidth]{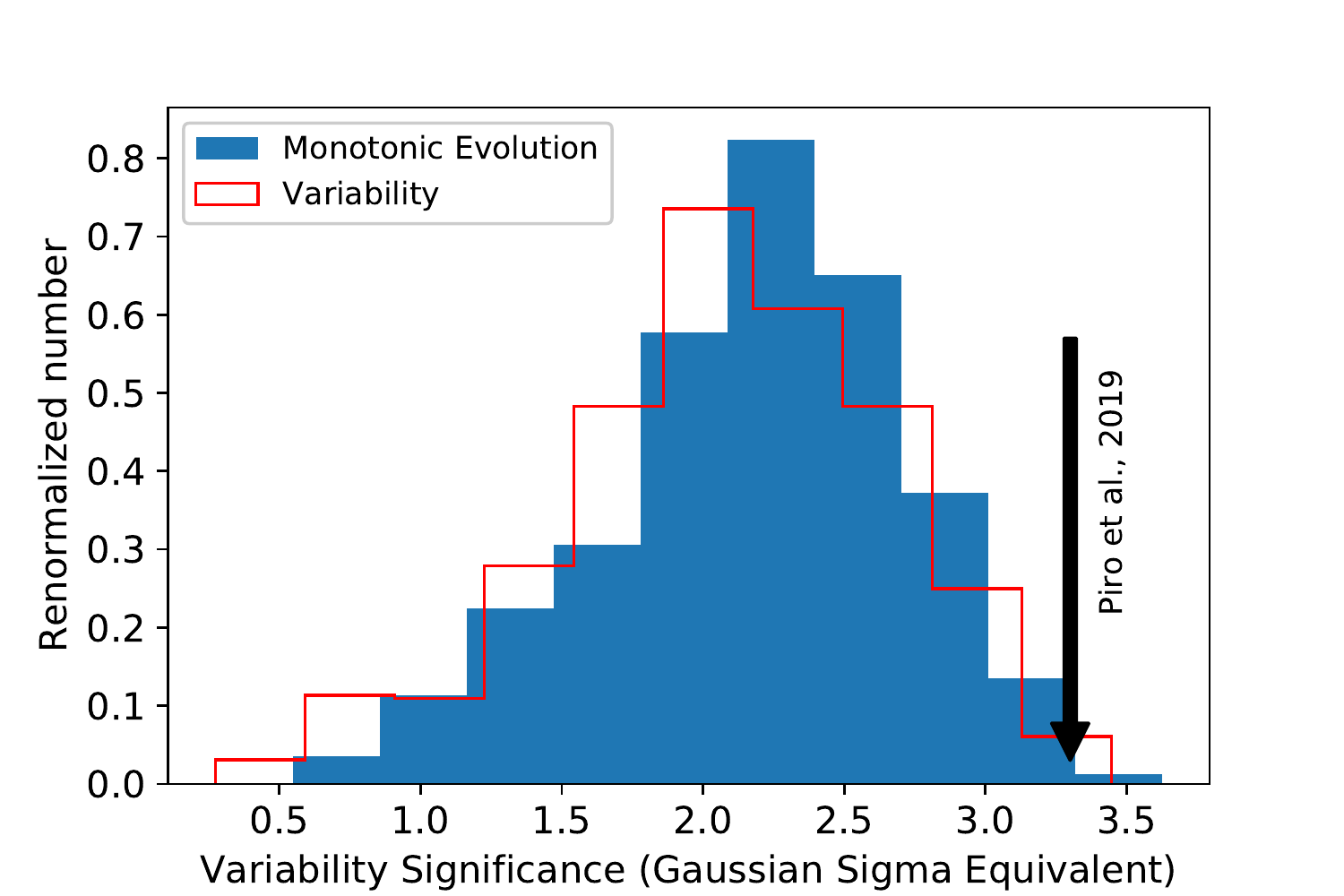}}
\caption{Distribution of statistical significance values (in units of Gaussian $\sigma$ equivalent) that quantifies the evidence for deviation from the $H_0$ hypothesis of a source with constant count-rate between $153-164$ days in the form of temporal variability (red), or monotonic evolution of the source count-rate (blue) for $10^{4}$ random selections of time intervals for comparison. A limited fraction of 0.3\% (red) or 0.4\% (blue) of randomly selected intervals show evidence for deviation from a constant count-rate at $\ge3.3\sigma$ c.l. The typical level of significance is $\sim2\,\sigma$.} 
\label{fig:variability}
\end{center}
\end{figure}

We carried out a blind search for deviations from a constant source count-rate in the time interval $\delta t=153-164$ days for which \cite{Piro+2019} report evidence for variability at the $3.3\,\sigma$ c.l. using two different approaches: (i) we divided the data set into two portions $\Delta t_1$ and  $\Delta t_2$, where the dividing line is randomly chosen within the $\Delta t$ of consideration, and we applied a Poissonian test to the number of detected photons $N_1$ and $N_2$. Our $H_0$ hypothesis is that $N_1$ and $N_2$ are randomly drawn from a Poisson distribution with expected rate $\lambda=1.49 \times 10^{-3} \rm{c\,s^{-1}}$ evaluated on the effective exposure times of the CXO during  $\Delta t_1$ and $\Delta t_2$ (i.e. the source count-rate is constant). We repeated the experiment $10^4$ times, considering only the cases with CXO exposure times during $\Delta t_1$ and $\Delta t_2$, $\Delta t_{1,exp}\ge \Delta t_{min}$ and $\Delta t_{2,exp}\ge \Delta t_{min}$, where $\Delta t_{min}=0.11$ d  is such that the probability of obtaining zero photons by chance is less than $P(\ge5\,\sigma)$ (i.e. $P(0)=e^{-\lambda \Delta t_{min}}<P(\ge5\,\sigma)$). The results from this exercise are shown in Fig. \ref{fig:variability}, red histogram. We find that a random selection of time intervals to compare typically leads to a $\sim2\,\sigma$ evidence for departure from our $H_0$ hypothesis of a constant count-rate, consistent with our results in Sec. \ref{SubSec:temp}, and that only $0.3\%$ of choices leads to a significance larger or equal to that  reported by \cite{Piro+2019}.  (ii) We further investigate the possibility of the presence of a monotonic evolution of the source count-rate, which would be best revealed by considering the initial and final portion of the data set only, as in \cite{Piro+2019}. We followed the same procedure as above and allowed for a random selection of the duration of the initial and final time intervals to consider within $\delta t=153-164$ days, with the constraint $\Delta t_{1,exp}\ge \Delta t_{min}$ and $\Delta t_{2,exp}\ge \Delta t_{min}$. Figure \ref{fig:variability} shows that only $0.4\%$ of the $10^{4}$ realizations that satisfy our constraints have evidence for a deviation from a constant count-rate with significance $\ge3.3\,\sigma$, and that the typical significance is $\sim2.2\,\sigma$. We conclude that the claim of a $3.3\,\sigma$ deviation from a constant count-rate by \cite{Piro+2019} mostly stems from comparing a particular selection of time intervals, and that a blind search for temporal variability on the same data set leads to a reduced statistical significance of $\sim2\,\sigma$.

\bibliography{ahbib2gw,ah_bib,radio}

\end{document}